\documentclass[10pt,twocolumn,twoside,submit]{JCNtran}
\usepackage{color}
\usepackage{amsmath}
\usepackage{amssymb}
\usepackage{tabularx}
\usepackage{epsfig}
\usepackage{graphicx}
\usepackage{epstopdf}
\usepackage{url}

\def\BibTeX{{\rm B\kern-.05em{\sc i\kern-.025em b}\kern-.08em
    T\kern-.1667em\lower.7ex\hbox{E}\kern-.125emX}}

\begin{document}
\bibliographystyle{jcn}
\title{DrivingStyles: A mobile platform for driving styles and fuel consumption
characterization }

\author{\noindent Javier E. Meseguer, C. K. Toh, Carlos T. Calafate, Juan
Carlos Cano, Pietro Manzoni}
\maketitle
\begin{abstract}
Intelligent Transportation Systems (ITS) rely on connected vehicle
applications to address real-world problems. Research is currently
being conducted to support safety, mobility and environmental applications.
This paper presents the DrivingStyles architecture, which adopts data
mining techniques and neural networks to analyze and generate a classification
of driving styles and fuel consumption based on driver characterization.
In particular, we have implemented an algorithm that is able to characterize
the degree of aggressiveness of each driver. We have also developed
a methodology to calculate, in real-time, the consumption and environmental
impact of spark ignition and diesel vehicles from a set of variables
obtained from the vehicle\textquoteright s Electronic Control Unit
(ECU). In this paper, we demonstrate the impact of the driving style
on fuel consumption, as well as its correlation with the greenhouse
gas emissions generated by each vehicle. Overall, our platform is
able to assist drivers in correcting their bad driving habits, while
offering helpful tips to improve fuel economy and driving safety.
\end{abstract}


\section{\label{sec:1 Introduction}Introduction}

Intelligent transportation systems (ITS) introduce advanced applications
aimed at providing innovative services, offering traffic management
and enabling users to be better informed, including support for safety,
mobility, and environmental applications. In parallel to ITS, mobile
devices have experienced technological breakthroughs in recent years,
evolving towards high performance terminals with multi-core microprocessors.
The smartphone is a clear representative outcome of this trend. 

In addition, the On Board Diagnostics (OBD-II) \cite{ISO14230} standard,
available since 1994, has recently become an enabling technology for
in-vehicle applications due to the availability of Bluetooth OBD-II
connectors. These connectors enable a transparent connectivity between
the mobile device and the vehicle's Electronic Control Unit (ECU).

When combining high performance smartphones with OBD-II connectivity,
new and exciting research challenges emerge, promoting the symbiosis
between vehicles and mobile devices, and thereby achieving novel intelligent
systems. DrivingStyles implements a solution based on neural networks,
which is capable of characterizing the driving style of each user
\cite{Meseguer}, as well as the fuel consumption \cite{MeseguerComsumption}.
In order to achieve this functionality, the data is obtained from
the ECU via the OBD-II Bluetooth interface, including the speed, acceleration,
revolutions per minute of the engine, mass flow sensor (MAF), manifold
absolute pressure (MAP), and intake air temperature (AIT). Currently,
this information can be collected and used in applications aimed at
improving road safety and promoting eco-driving, thus reducing fuel
consumption and greenhouse gas emissions. Specifically we find that,
by shifting towards a more efficient driving style, users can save
up to 20\% of fuel while improving driving safety, thereby reducing
greenhouse gases as we detail later on.

This paper is organized as follows: in the next section we present
some related works. Section \ref{sec:DrivingStyles Architecture}
introduces the DrivingStyles architecture (both the Android and the
server interface). Models for fuel consumption and $CO_{2}$ emissions,
are described in more detail in section \ref{sec:Fuel-Consumption-Calculation}.
The tuning of the neural network, along with the obtained results,
are presented in sections \ref{sec:Neural Networks} and \ref{sec:Experimental results},
respectively. Finally, section \ref{sec:Conclusions-and-future} presents
the conclusion of our work.

\begin{figure*}[t]
\begin{centering}
\includegraphics[width=0.9\linewidth]{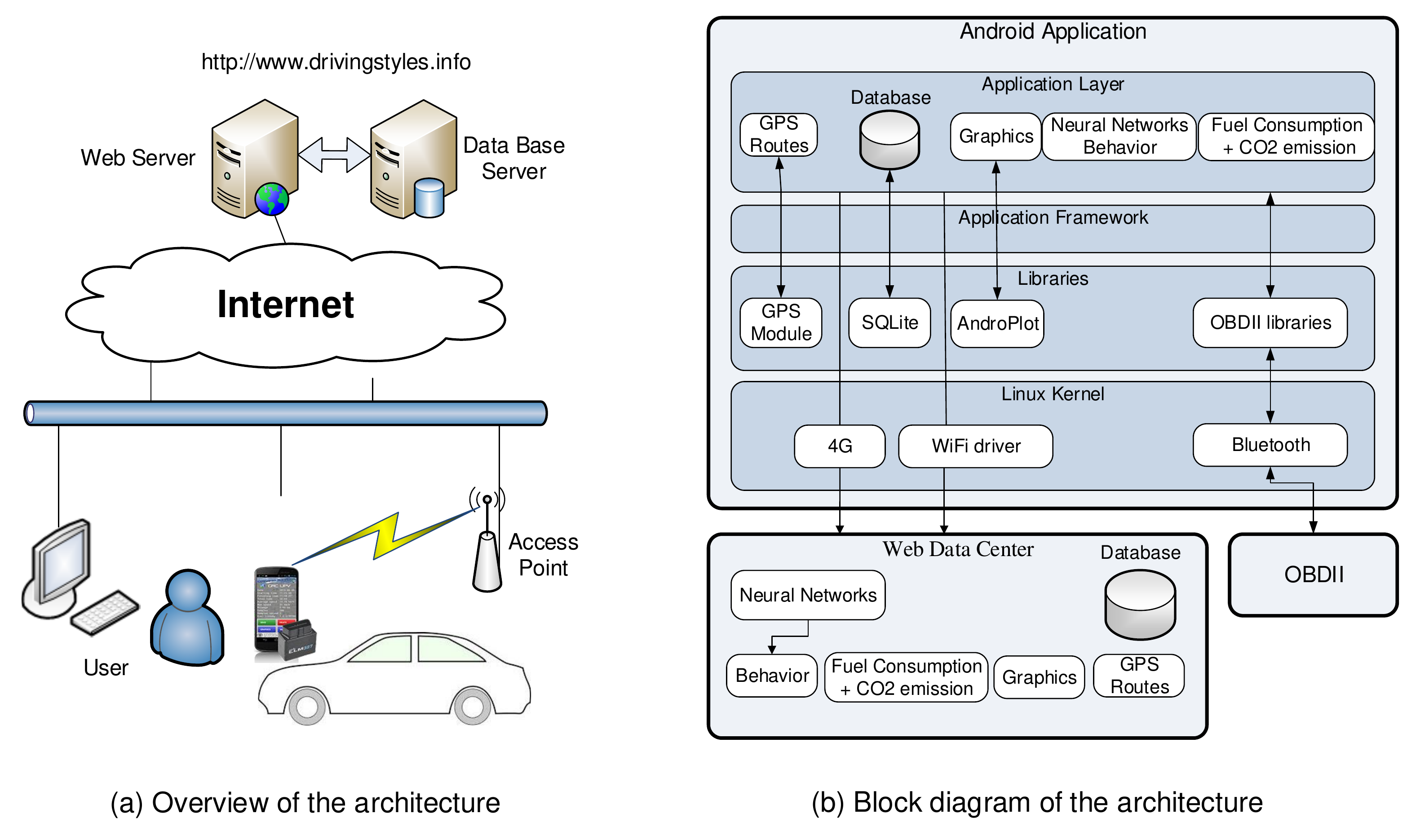}
\par\end{centering}
\caption{\label{fig:Architecture}System Architecture of DrivingStyles.}
\end{figure*}

\section{\label{sec:Related-Work}Related Work}

Technological advancements in the field of mobile telephony are making
smartphones very powerful. This high computing power opens new and
attractive opportunities for research. When coupled with the eco-driving
concept, it has gained great significance in recent years \cite{Somayajula}.
An example is the prototype of an onboard unit developed by Hernandez
et al. \cite{Hernandez}. These driving techniques save fuel consumption,
regardless of the technology used inside the vehicle. One of the main
problems of eco-driving systems is identifying the factors that affect
energy consumption. Ericsson \cite{Erikcsson} suggests that, in order
to save fuel, sudden changes in acceleration and high speed driving
should be avoided. Johansson et al. \cite{Johansson} suggest maintaining
low levels of deceleration, minimizing the use of the first and second
gears, and putting every effort into using the 5th and 6th gears,
while avoiding continuous gear changes.

There are several proposals that analyze which variables affect fuel
consumption. Kuhler \cite{Kuhler} introduced a set of ten variables
that are used in laboratories for fuel consumption and vehicle emissions
analysis. Other authors such as André \cite{Andre} improve these
results by increasing and replacing some of the parameters. In previous
works such as Leung \cite{DYC} and COPERT III \cite{COPERTIII},
different tools were developed to enable real-time collection of engine
and vehicle parameters from the OBD connector. 

Several commercial OBD-II scanner tools are available that can read
and record these sensor values. Apart from such scanners, remote diagnostic
systems such as GM\textquoteright s OnStar, BMW\textquoteright s Connected
Drive, and Lexus Link are capable of monitoring engine parameters
from a remote location. Car manufacturers used eco-monitoring to reflect
the instant, historical, and time-elapsed fuel economy, and is used
in the car through on-board trip computers \cite{Matsumoto}. Our
solution differs from all the previous ones by providing a real-time
analysis of the driving style of each user in the scope of eco-driving
behavior, and based on neural network techniques. By calculating the
consumption and greenhouse gas emissions generated by both types of
engines (spark ignition, and diesel vehicles), we are able to closely
relate both results, detailing the fuel savings achieved by soft driving
patterns when compared to aggressive ones. 

\section{\label{sec:DrivingStyles Architecture}DrivingStyles Architecture}

Our proposed architecture applies data mining techniques to generate
a classification of the driving styles of users based on the analysis
of their mobility traces. Such classification is generated taking
into consideration the characteristics of each route, such as whether
it is urban, suburban, or a highway, and it is then correlated with
the fuel consumption and emissions of each driver.

To achieve the overall objective, our system comprises four elements:
\begin{enumerate}
\item An application for Android, based smartphones. Using an OBD-II Bluetooth
interface, the application collects control information (by default
every second, but it is configurable by the user) such as speed, acceleration,
engine revolutions per minute, throttle position, and the vehicle's
geographic position. In addition, we also obtain via OBD-II the mass
flow sensor (MAF), the manifold absolute pressure (MAP), and the intake
air temperature (AIT) that are used in the calculation of fuel consumption.
After gathering the information, the user can upload the collected
data to the remote data center for analysis.
\item A data center offering a web interface to collect large data sets
sent by different users concurrently, and to graphically display a
summary of the most relevant results, like driving styles and route
characterization of each route sent. Our solution is based on open
source software tools such as Apache, PHP and Joomla. 
\item A neural network, which has been trained using the most representative
route traces in order to correctly identify, for each path segment,
the driving style of the driver, as well as the segment profile: urban,
suburban or highway. We use the backpropagation algorithm \cite{Hecht-Nielsen},
which has proven to provide good results in classification problems
such as the one associated to this project.
\item Integration of the tuned neural networks both within the mobile device
itself, and in the data-center platform. The goal is to use neural
networks to dynamically and automatically analyze user data, reporting
to the drivers in real time and allowing them to find out their driver
profile, thus promoting a less aggressive and more ecological driving.
\end{enumerate}
The block diagram of the DrivingStyles architecture is shown in figure
\ref{fig:Architecture}b. This consists of three blocks: the mobile
application on a Android device, the data center platform, and an
On Board Diagnostics (OBD-II) device. 

The basic layer in the Android device is the Linux kernel, which contains
all the essential hardware drivers to interact with the OBD-II device
via Bluetooth. The top layer includes both Android's
native libraries and our own libraries. Specifically, we developed
the OBD-II communications module, along with the libraries for graphical
data representation, at this layer. The next level up is the Application
Framework; this layer manages the basic functions of the mobile device,
and the communications with the developed libraries. 

Finally, at the application layer, we developed the different modules
of the DrivingStyles architecture, such as the fuel consumption and
$CO_{2}$ emissions estimators, the neural networks behavior, GPS
routes, and graphics. Also, the application provides real-time feedback
from the device to the user such that, when it detects high levels
of aggressiveness (above a certain threshold), the device automatically
generates an acoustic signal to alert the driver. 

\subsection{\label{sec:Android application}DrivingStyles Android Interface}

The first step for a user is to register at \url{http://www.drivingstyles.info},
and to download the free Android application. After installing the
Android application in the mobile device, and after connecting to
the Bluetooth ELM327 interface inside the car (this connector is mandatory
on all vehicles since 2001), the data acquisition process will start
(see figure \ref{fig:Architecture}a).

The Android application is a key element of our system, proving connectivity
to the vehicle and to the DrivingStyles web platform. Currently, it
can be downloaded for free from the DrivingStyles website\footnote{\url{http://www.drivingstyles.info}},
or from Google Play \footnote{\url{https://play.google.com/store/apps/details?id=com.driving.styles}}.

Once the mobile application is installed and configured, the user
must pair the mobile device with the ELM327 (OBD-II Bluetooth device)
to start getting data. The data obtained from the different variables
such as acceleration, engine revolutions per minute (RPM), speed,
mass flow sensor (MAF), manifold absolute pressure (MAP), and intake
air temperature (IAT) are analyzed by the application, showing users
the characteristics related to their driving, fuel consumption, and
$CO_{2}$ emissions. 

\begin{figure}[t]
\begin{centering}
\includegraphics[width=1\linewidth]{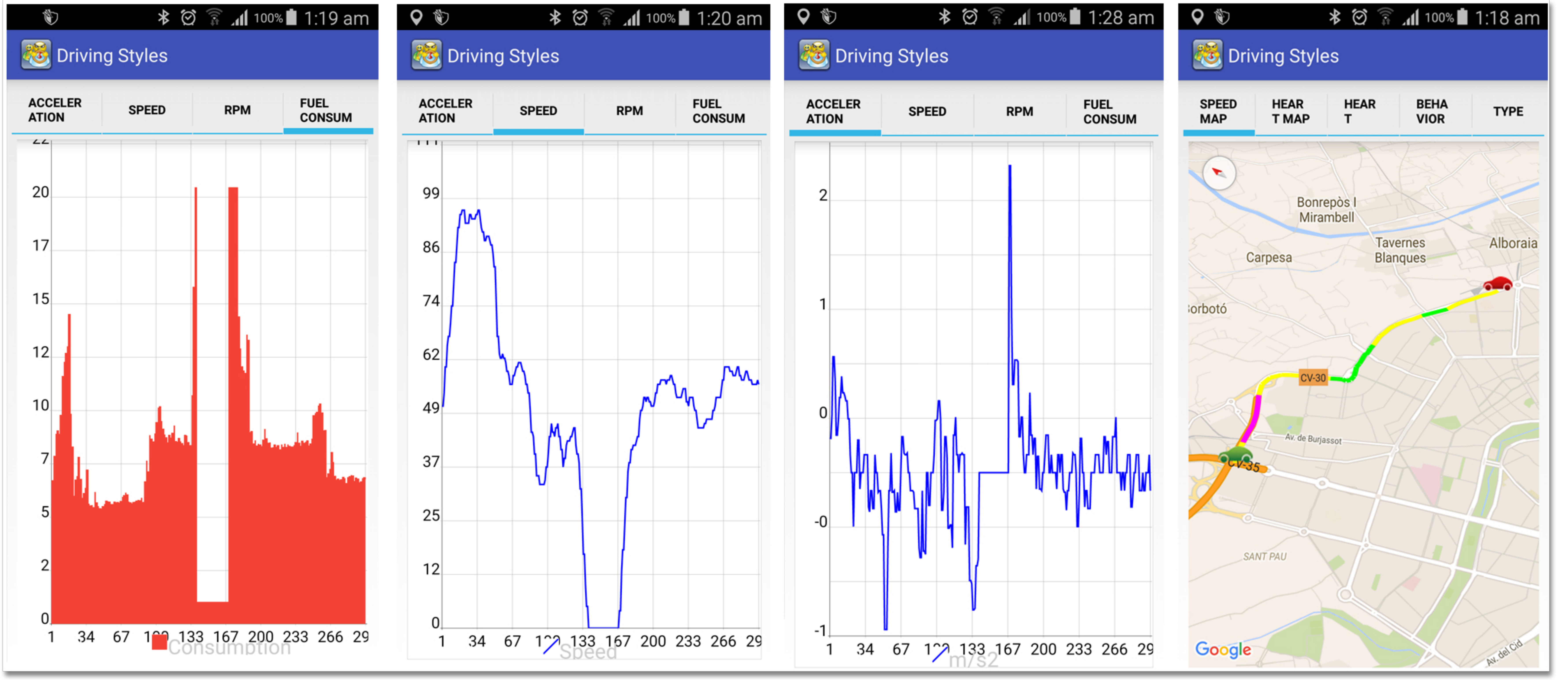}\caption{Snapshots of the fuel consumption, speed, acceleration parameters,
and map module for the Android application.}
\par\end{centering}
\end{figure}

In order to adjust the application functionality, it offers several
configuration options, i.e., User creation, Connection options, GPS
Activation, Sensor sampling, and type of fuel definition. Once configured,
our application captures data sent by the OBD-II and the GPS interfaces,
as well as the phone\textquoteright s accelerometer showing the monitored
sensors, and performing several monitoring actions in real time without
affecting the data captured. Figure 2, shows some snapshots of our
DrivingStyles application.

In addition, routes traces can also be sent to the website data center
for further analysis. This module can be accessed either from the
historic stored routes, or immediately after stopping the data capture.
The information screen displays the header information of the selected
route, such as: (i) date of the captured data, (ii) start time, (iii)
finish time, (iv) maximum speed, and (v) fuel consumption. The URL
of the DrivingStyles web interface is http://www.drivingstyles.info.

\subsection{\label{sec:Application WEB: DrivingStyles}DrivingStyles Server Interface}

\begin{figure*}[t]
\begin{centering}
\includegraphics[width=0.8\textwidth]{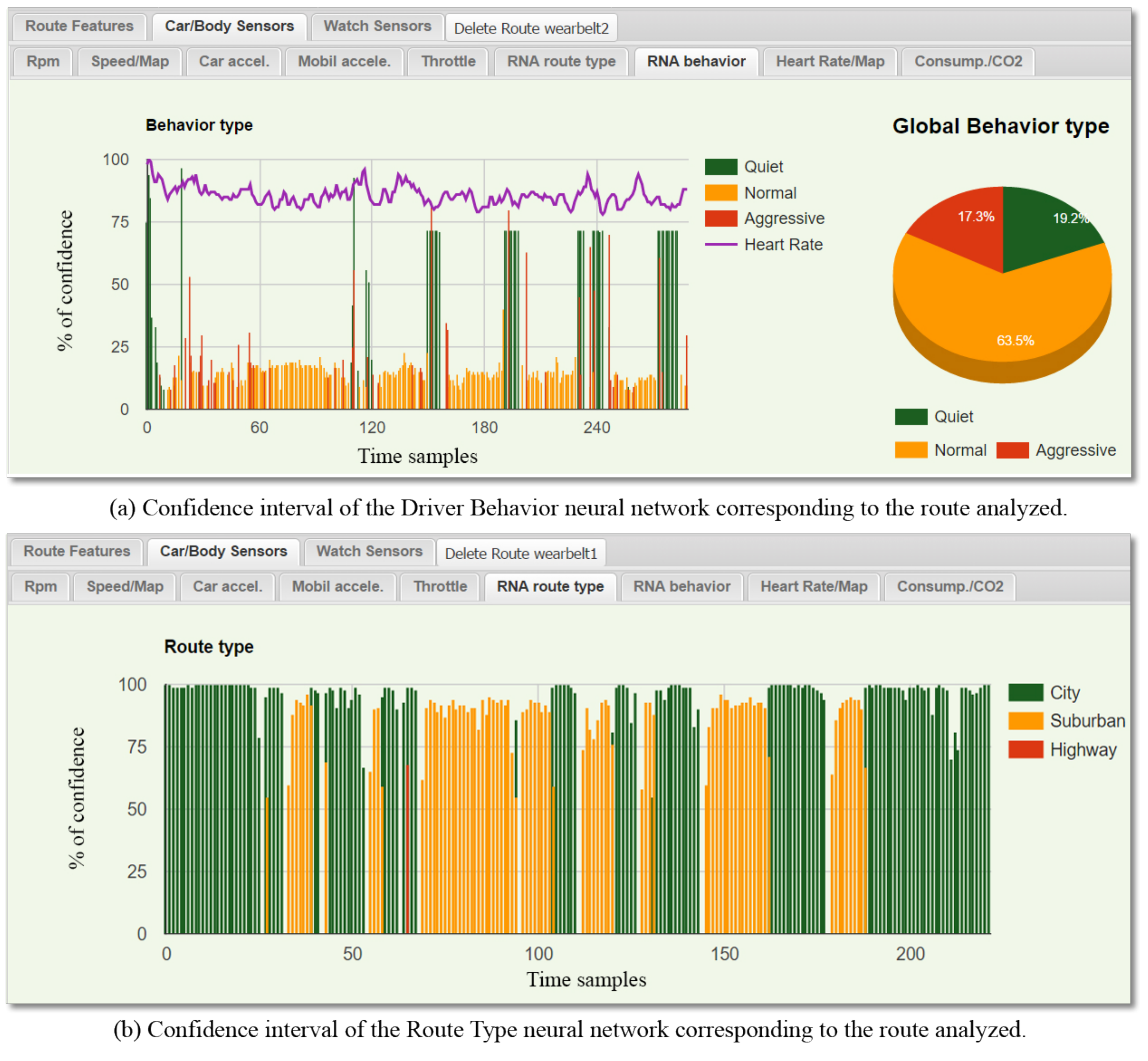}
\par\end{centering}
\centering{}\caption{\label{fig:Snapshots_RNA_graphics}Snapshots of driver behavior and
route type corresponding to the data center web.}
\end{figure*}

The second main component of our architecture corresponds to the data
center and its web interface. To implement this component, we have
selected open source software such as Apache HTTP, and Joomla as the
content management system (CMS). We have used a CMS, combined with
the use of a resource wrapper, which detachs our system from the presentation
layer, thus focusing on the driving styles characterization problem.
This module can be found in \url{http://www.drivingstyles.info}.

Basically the server receives data sent from the Android application
of each user, and it provides functionality to work with User, Routes,
and Statistics.

Once the user is logged in, he is asked to record a number of important
data, especially for future data mining studies. The most relevant
items are sex, age, and other details concerning the vehicle used:
car manufacturer, model, fuel type, and the theoretical 0-100 acceleration
level (important to normalize the user behavior in our study). 

In the Routes' section, users can access all the routes they have
uploaded. When selecting car/body sensors, the system displays nine
graphs for the different sensors obtained from the OBD-II (direct
and indirect variables), as well as the route and driver behavior
(see figure \ref{fig:Snapshots_RNA_graphics}). Next, we present our
fuel consumption estimation approach relating it with the driver style
as captured by the DrivingStyles platform. 

\section{\label{sec:Fuel-Consumption-Calculation}Fuel consumption and greenhouse
gas emissions calculation}

\subsection{Fuel consumption}

Fuel consumption is usually represented as the ratio of fuel consumed
per distance travelled, being measured in terms of litres per 100
kilometres (or alternatively as MPG - miles per gallon).

In this work, we focus on gasoline and diesel engines. Although the
basic designs of gasoline and diesel engines are similar, the mechanics
are different. 

A gasoline engine compresses its fuel and air charge and then initiates
combustion by the use of a spark plug. A diesel engine just compresses
air until the combustion chamber reaches a temperature for self-ignition
to occur. So, at a given speed in kilometres per hour, instantaneous
fuel consumption can be calculated as follows:

\begin{equation}
Inst.\:Fuel\:Consump.\left[l/km\right]=\frac{Fuel\:Flow\:\left[l\right]}{Speed\:\left[km\right]}\label{eq:InstantaneousFuelConsumption}
\end{equation}

\noindent \begin{flushleft}
Notice that it can only be calculated when the vehicle is moving and
the engine is operating.
\par\end{flushleft}

In addition, the Fuel Flow PID must be available, which often does
not occur since most vehicles fail to support all the standard OBD
PIDs. In fact, although there are many manufacturer-defined custom
PIDs (not part of the OBD-II standard), the OBD standard itself does
not provide a fuel consumption parameter. Instead, it provides other
values that enable its calculation. Depending on the variables that
the ECU can supply, the mathematical procedure to derive fuel consumption
is different, as described below (see figure \ref{fig:MAF calculation})
:
\begin{enumerate}
\item By combining the Engine Fuel Rate (PID 015E), also known as Fuel Flow
(litres/hour), and Speed (PID  010D), it is easy to calculate instantaneous
fuel consumption. However, while speed is mandatorily available, fuel
rate is not. In fact, it was unavailable in all vehicles we used to
carry out our tests. This can be due to two reasons: (i) the manufacturer
chooses not to make it available, or (ii) there is no sensor inserted
in the fuel line between the fuel tank and the engine carburetor to
measure litres per hour.
\item If the MAF PID is available, but the Engine Fuel Rate is not, we can
calculate fuel rate as Fuel Flow (litres/hour) by dividing the Mass
Air Flow (PID 0110) $\:\cdot\:$ 3600 sec. by the product of air-to-fuel
ratio and Fuel Density (using a fuel density equal to 820 g/dm3 for
gasoline and 720 g/dm3 for diesel): 
\begin{equation}
Fuel\:Flow\,\left[l/h\right]=(MAF\cdot3600)/AFR_{A}\cdot FD\label{eq:FuelFlow}
\end{equation}
 where $MAF$ refers to Mass Air Flow (g/s), $AFR_{A}$ to the actual
Air-to-Fuel Ratio (being 14.7 and 14.5 grams of air to 1 gram of fuel
for gasoline and diesel respectively), and $FD$ to the Fuel Density.
The ratio between Fuel Flow and Speed, allows us to directly calculate
fuel consumption.
\item Finally, If MAF is not available , there are two additional ways to
calculate it (See \cite{Bruce} for more details): 
\end{enumerate}
\begin{itemize}
\item As a function of the absolute load (PID 0143), the RPM (PID 010C)
and the Engine Displacement (EngDisp, volume of an engine's cylinders
in $cm^{3}$), intake stroke is the fluid admission phase of a reciprocating
cylinder.
\item As a function of the intake manifold pressure (PID 010B), RPM (PID
010C), intake air temperature (PID 010F), and engine displacement.
\begin{figure}[t]
\begin{centering}
\includegraphics[width=1\columnwidth]{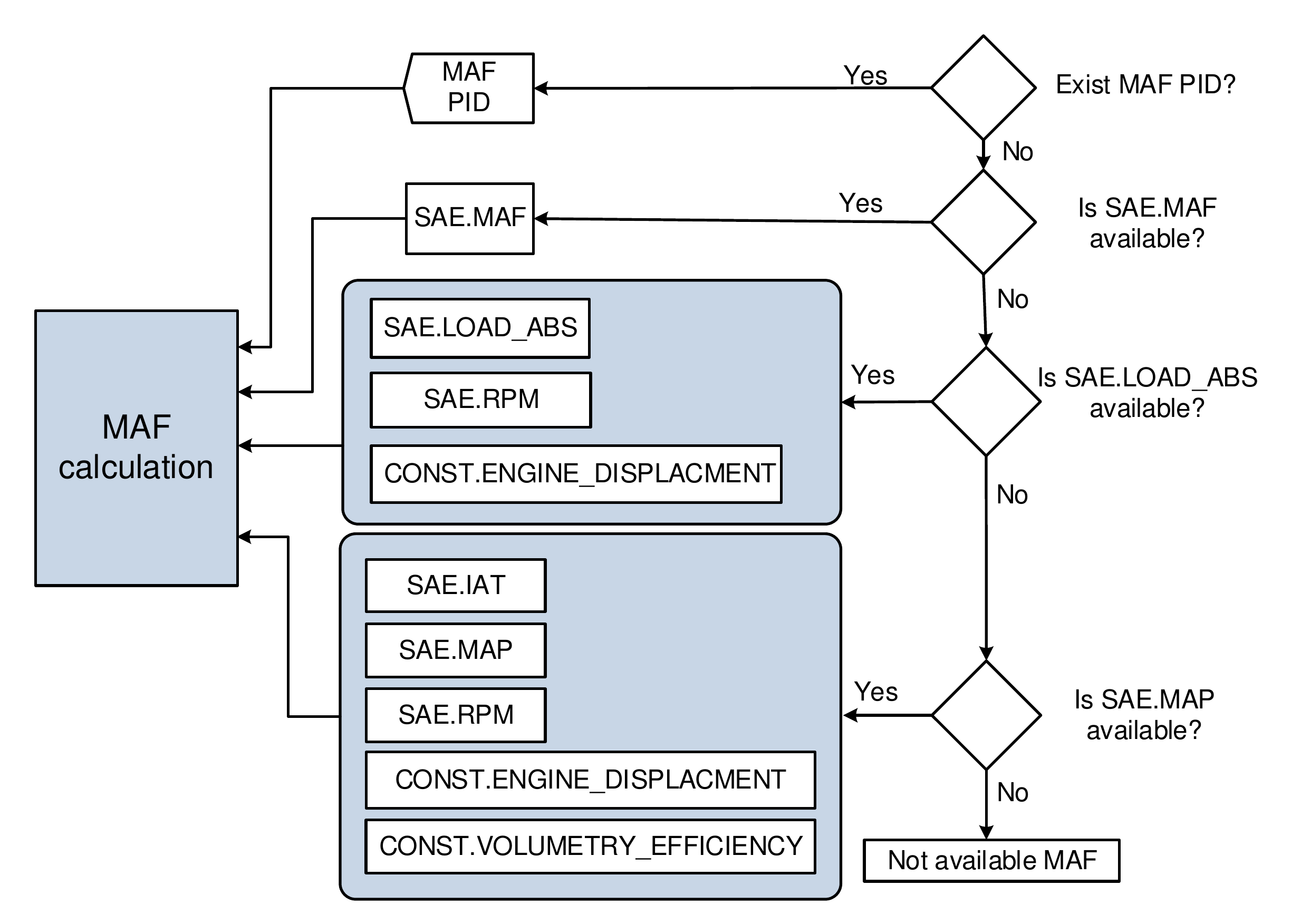}
\par\end{centering}
\caption{\label{fig:MAF calculation}Scheme of the different MAF calculation
possibilities regarding fuel consumption calculation.}

\end{figure}
\end{itemize}
\begin{figure*}[t]
\begin{centering}
\includegraphics[width=1\textwidth]{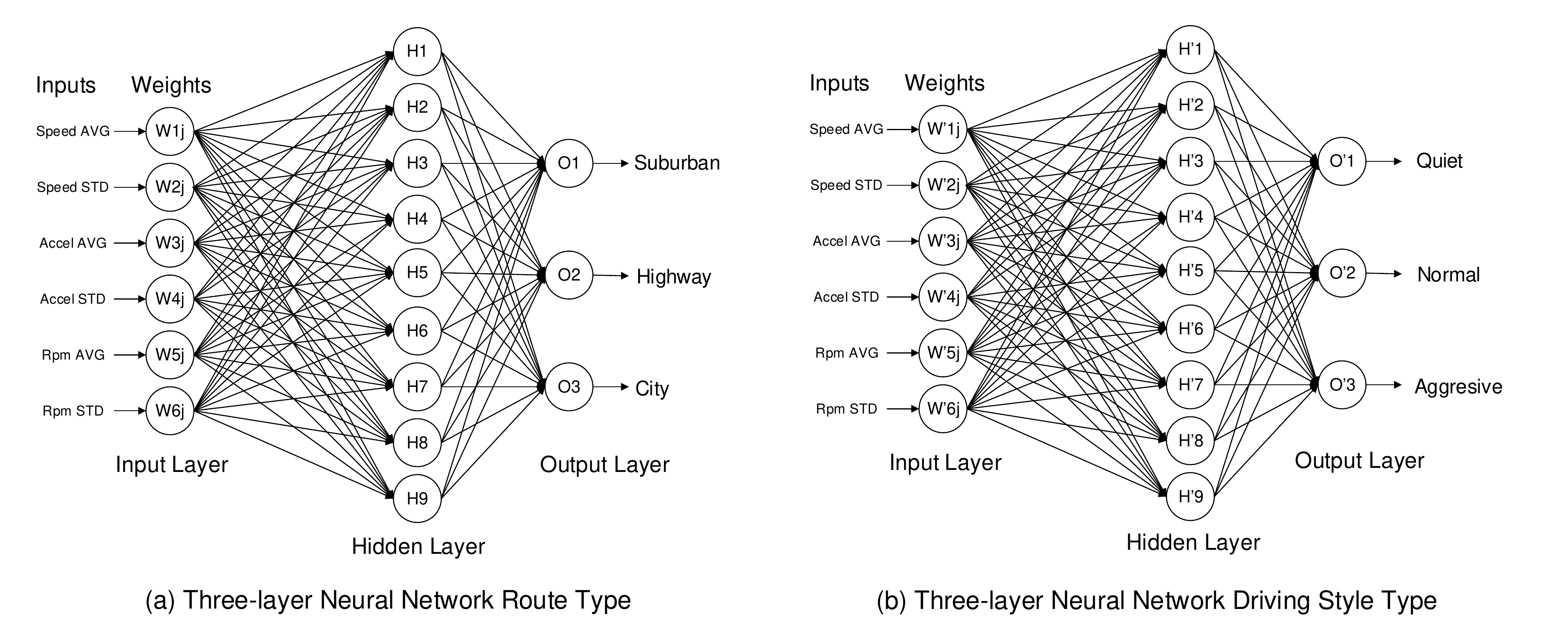}
\par\end{centering}
\caption{\label{fig:Schematic Representation of the Three-layer Neural Networks usedused}Schematic
representation of the Three-layer Neural Networks used in the DrivingStyles
Architecture. }
\end{figure*}

\subsection{Greenhouse gas emissions calculation}

The most significant greenhouse gases are generated from direct combustion
of carbon dioxide $CO_{2}$, Methane ($CH_{4}$), and Nitrous oxide
($N_{2}O$), among others. $CO_{2}$ is always generated when burning
fuel that contains carbon. Since the carbon in the fuel is combined
with the oxygen in the air: $C+O_{2}\rightarrow CO_{2}$, the amount
of $CO_{2}$ can be calculated by the atomic masses of carbon and
oxygen, and the carbon content of the fuel. The atomic mass of carbon
is $12_{U}$ and oxygen is $16_{U}$, meaning that $CO_{2}=12_{U}+2\:\cdot\:16_{U}=44_{U}$.
Burning 1kg of carbon produces $44/12\approx3.67kg$ of $CO_{2}$
in complete combustion, and so the $CO_{2}$ emission of combustion
is $3.67\:\cdot\:C_{c}\:\cdot\:m_{fuel}$ where $C_{c}=fuel\:carbon\:content\:(mass\:basis)$.
Considering that the carbon content of diesel fuel is 85.7\% the $CO_{2}$
emission when burning 1kg ($m_{fuel}=1kg$) of diesel fuel is:

\begin{eqnarray}
m_{CO_{2}} & = & 3.67\:\cdot\:C_{c}\:\cdot\:m_{fuel}\nonumber \\
m_{CO_{2}} & = & 3.67\:\cdot\:0.857\:\cdot\:1\left[kg\right]=3.15\:\left[kg/1kg\:fuel\right]\nonumber \\
 &  & Density\:of\:diesel\:fuel\:is\:0.84\:\left[kg/l\right]\nonumber \\
m_{CO_{2}} & = & 3.15\:\cdot{\left[kg\right]}\:\cdot\:0.84=2.64\:\left[kg/1l\:fuel\right]\label{eq:mCO2}
\end{eqnarray}

Driving in a fuel-efficient manner can save fuel, money, and reduce
greenhouse gas emissions. Among the factors that can affect fuel consumption,
such as: vehicle age and condition, outside temperature, weather,
and traffic conditions, we consider that driver behavior can be one
of the most relevant parameter. Next, we provide detailed information
about the neural network we proposed for characterizing driver styles.

\section{\label{sec:Neural Networks}Neural Networks-based data analysis}

Neural networks \cite{Haykin} use artificial intelligence and automatic
processing techniques to learn how to find patterns in data, thereby
improving their success rate at making decisions of predictions. A
learning algorithm is used to generate the neural network. For example,
the driving style of each user and the type of route can be characterized
from a well-defined set of rules and the ECU input variables.

There are many different learning algorithms such as backprop\_momentum,
Hebbian, or delta-rule, each one having its own advantages and disadvantages
depending on the type of problems. In our project, we face a classification
problem: starting from some input data, which in our case are the
speed, acceleration, and revolutions per minute (rpm) of the engine,
we intend to obtain as outputs the type of road and the driving style.
The problem of classifying the driver behavior and the route type
with a supervised learning is to find a function that best maps a
set of inputs to its correct output. We tried several types of algorithms
in this direction, including backpropagation, backprop-momentum, and
batch backpropagation, and the results evidenced that backpropagation
\cite{Hecht-Nielsen,Classification} was the best algorithm for our
study since it achieved the lowest sum of squared errors (SSE) in
terms of prediction.

A data preprocessing stage is selected from all the possible input
variables of the neural network. From all the possible data, we keep
a subset of these variables. In practice, this subset is not the minimum;
instead, it is a compromise between a manageable number (not too large)
of variables and an acceptable network performance. In this work,
after considering the many variables that can be obtained from the
Electronic Control Unit (ECU), we have chosen to train the neural
network using: the mean and standard deviation of speed, the vehicle
acceleration, and the rpm value. 

In all the vehicles used for testing, these variables are easily obtained.
Other variables, such as the position of the throttle, which would
provide important information for the neural network training, have
to be rejected because not all manufacturers provide such information.
The data input of each parameter is normalized between 0 and 1; this
normalization should take into consideration the range of possible
values. The schematic representation of our three-layer neural network
can be seen in figures \ref{fig:Schematic Representation of the Three-layer Neural Networks usedused}a,
and \ref{fig:Schematic Representation of the Three-layer Neural Networks usedused}b.

The application used for the creation and training of the neural networks
required by this project is JavaNNS \cite{JavaNeuralNetworkSimulator},
a java version of the SNNS program from the University of Tübingen.

First, an empty neural network was created, defining the number of
entries mentioned previously, and the number of hidden nodes. A larger
number of hidden nodes can improve the success rate, but has the negative
effect of increasing the response time. On the contrary, with a large
number of nodes, the network becomes a memory bank that can recall
the training set to perfection, but does not perform well on samples
that were not part of the training set. There are three output nodes
for each neural network, one that characterizes the type of road (urban,
suburban, or highway), and another one that characterizes the user's
driving style (quiet, normal, or aggressive), see figures \ref{fig:Architecture}a
and \ref{fig:Architecture}b respectively.

Subsequently, we train the network with a total of 16038 samples,
each representing a 3-second drive period (13.3 hours in total belonging
to 7 drivers of different ages and sex). We initially adjust the learning
rate to learning intervals of 0.2, and then modify this value to observe
how the error affects the neural network (JavaNNS application \cite{JavaNeuralNetworkSimulator}
computes the mean square error in each learning iteration). The higher
the learning rate, the greater the weight updating following each
iteration; therefore, learning becomes faster, but it is prone to
cause unwanted oscillations in the network. As the network training
progresses, the number of learning cycles that take place in the tests
is adjusted until the final trained network is obtained.

Once the neural network was successfully trained, the knowledge obtained
was converted into C code, and this code was then integrated into
our DrivingStyles platform. With the neural network already implemented,
every time a route or route segment is selected, the system automatically
returns the type of road, as well as the associated driving style.
Figure \ref{fig:Snapshots_RNA_graphics} shows the results obtained
by our neural network including the driving styles and route characterization
of a particular route of one of our users. 

Overall, with different traces analyzed, along with the drivers using
our application, we have shown a correct classification of the different
routes registered, both in terms of route types and driving styles,
thus validating our proposed solution.

\section{\label{sec:Experimental results}Experimental results and evaluation}

\begin{figure}[t]
\begin{centering}
\includegraphics[width=1\columnwidth]{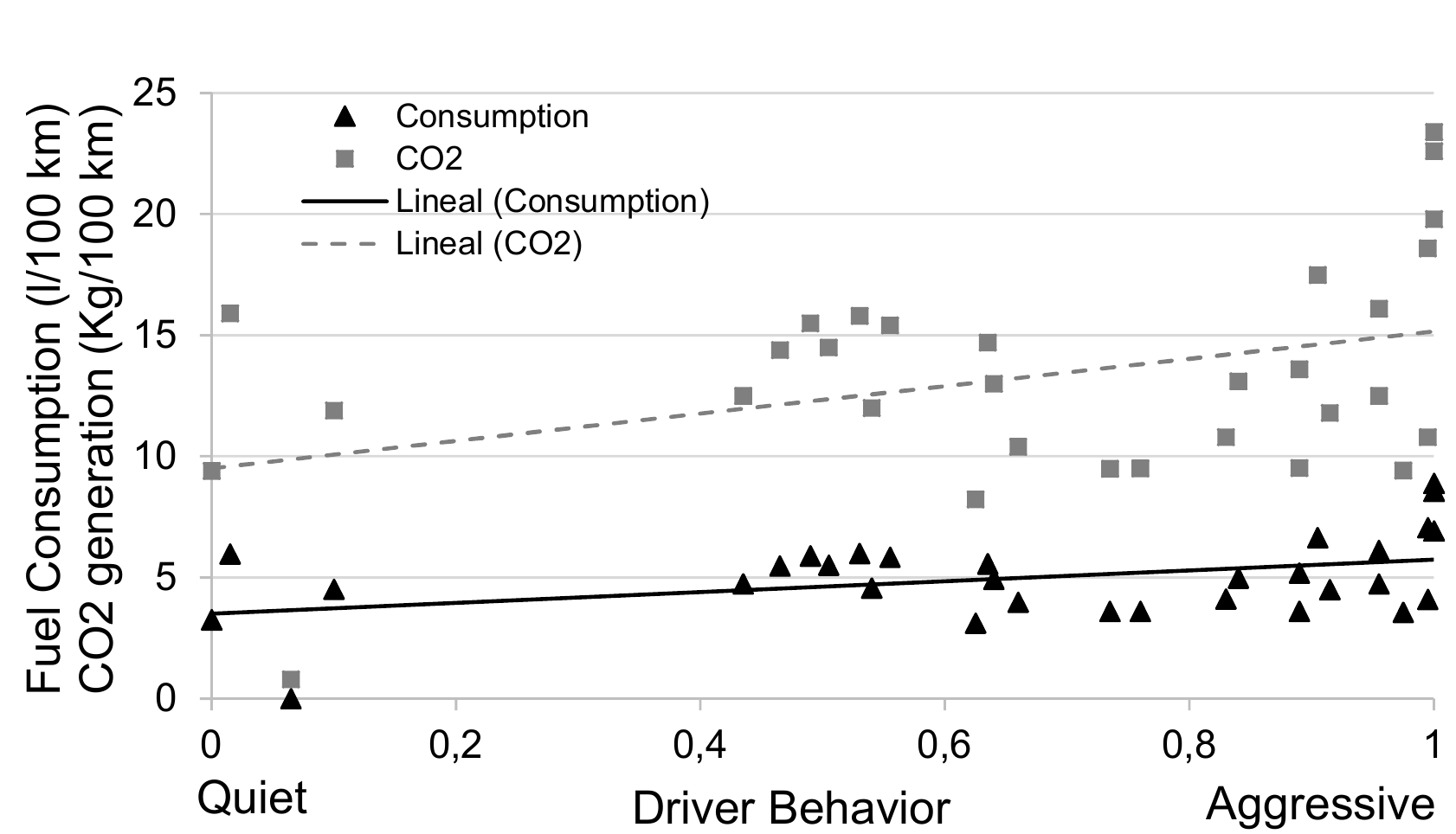}
\par\end{centering}
\caption{\label{fig:Chart-of-consumption}Chart of consumption and $CO_{2}$
in relation to the driving behavior.}
\end{figure}

In our project, we focus on characterizing the driving style of different
drivers, and then measuring the associated fuel consumption variations.
In order to achieve this objective, we rely on the collaboration of
534 drivers from around the world using our platform, including countries
like India, Brazil, Central America, and Europe. In this particular
study, we analyzed the behavior of 75 representative routes (each
divided into 10 second periods) using the neural network described
earlier. For each section, the neural network returns the corresponding
driver behavior, and we combine this data with the fuel consumption
data corresponding to that route.

\begin{figure}[H]
\begin{centering}
\includegraphics[width=1\columnwidth]{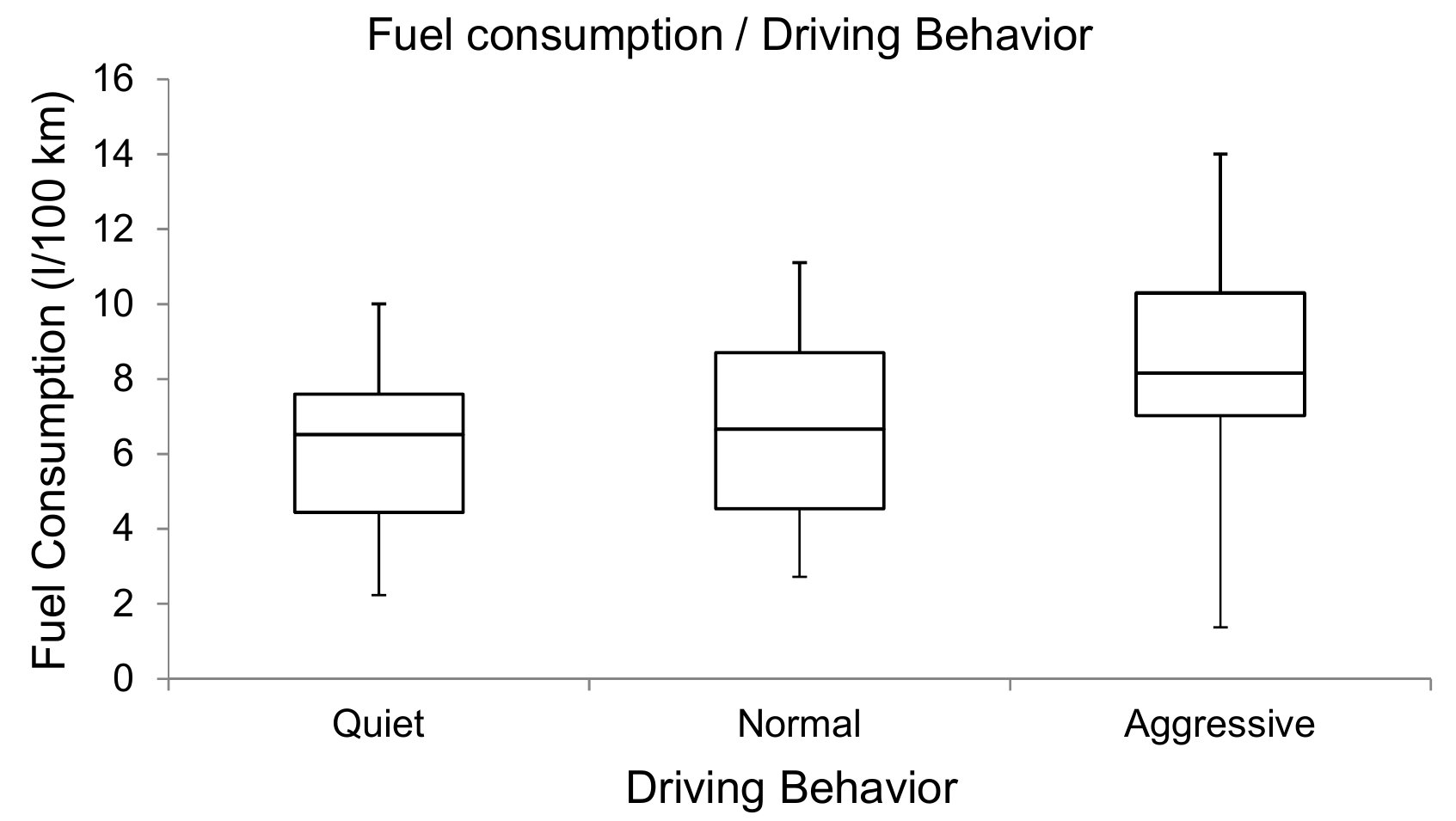}
\par\end{centering}
\caption{\label{fig:Box-and-whisker}Box and whisker plot of Fuel Consumption/Driving
Behavior.}
\end{figure}

We carried out several types of tests to validate our proposals. Figure
\ref{fig:Chart-of-consumption} shows the fuel consumption and $CO_{2}$
emissions reported by different drivers classified according to their
driving style. The results of this test show that a more aggressive
driving behavior causes fuel consumption to increase significantly,
while also increasing the generation of $CO_{2}$. To gain further
insight into these correlations, Figure \ref{fig:Box-and-whisker}
displays the differences between quiet, normal, and aggressive driving
behavior in terms of fuel consumption; aggressive drivers provoke
fast starts and quick accelerations, driving at high engine revolutions,
and causing sudden speed changes. Conversely, a quiet driving behavior
would be smooth, without sudden speed changes or continuous gear shifts.
It is clear that fuel consumption increases when the driver behavior
becomes more aggressive, with average differences of up to 1.5 liters
per 100km. In our experiments, an aggressive driver uses an average
of 8 liters per 100km, and a quiet driver only 6.6 liters per 100km,
meaning that the difference in terms of fuel consumption is not negligible,
as the former may consume up to 20\% more fuel depending on the driving
style. Regarding $CO_{2}$ emissions, they may increase by 50\%, going
from 10 to 15Kg/100km, depending on whether drivers are quiet or aggressive.

\section{\label{sec:Conclusions-and-future}Conclusions and future work}

This paper presents our DrivingStyles platform, which integrates mobile
devices with data obtained from the vehicle's engine Electronic Control
Unit (ECU) to characterize driver habits, as well as the associated
fuel consumption and emissions. Our platform helps to promote a more
ecological driving style by emphasizing on the relationship between
driving style and fuel consumption, which has a clear and direct impact
on the environment. It has
been also demonstrated that the driving style is directly related
to fuel consumption. Specifically, adopting an efficient driving style
allows achieving fuel savings ranging from 15 to 20\%. An aggressive
driving style always results in a greater energy consumption and more
$CO_{2}$ emissions, whereas smooth driving ends up providing a greater
energy efficiency and reduced gas emissions. The application, which
is available for free download in the DrivingStyle\textquoteright s
website and Google Play Store, achieved more than 5800 downloads from
different countries in just a few months. This emphasizes the great
interest about research on this topic. As future work, we intend to
extend this platform by providing route recommendations based on real-time
feedback about the congestion state of different alternative routes,
as well as providing estimated greenhouse emissions for different
routes.

\section*{Acknowledgments}

This work was partially supported by the Ministerio de Economía y
Competitividad, Programa Estatal de Investigación, Desarrollo e Innovación
Orientada a los Retos de la Sociedad, Proyectos I+D+I 2014, Spain,
under Grant TEC2014-52690-R.

\vspace{5pt}
\bibliographystyle{IEEE}

\epsfysize=3.2cm
\begin{biography}{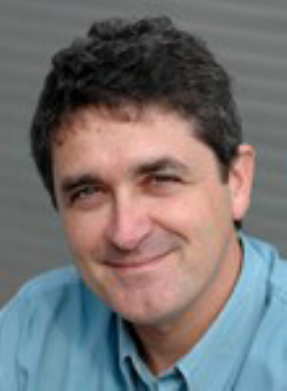}{Javier E. Meseguer}(jmesegue@upv.es) is a researcher in the Institute for Molecular
Imaging Instrumentation (I3M). He has been involved in Web, Cloud
development and Grid Technologies in the past 12 years, participating
in 14 National and European Research Projects. From 1999 to 2004 he
worked as a consultant at PricewaterhouseCooper and IBM Global services
in Madrid.
\end{biography} 
 
\epsfysize=3.2cm
\begin{biography}{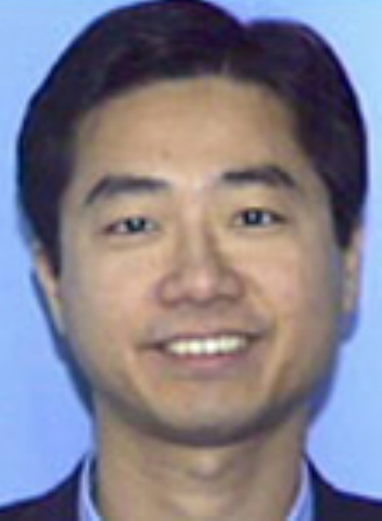}{C. K. Toh} received his Ph.D. degree in Computer Science from Cambridge University
(1996) and earlier EE degrees from Manchester University (1991) and
Singapore Polytechnic (1986). TRWTactical Systems Inc., USA.
DARPA Deployable Ad Hoc Networks Program at Hughes Research Labs,
USA. He is a Fellow of the British
Computer Society (BCS), New Zealand Computer Society (NZCS), Hong
Kong Institution of Engineers (HKIE), and Institution of Electrical
Engineers (IEE). He is a recipient of the 2005 IEEE Institution Kiyo
Tomiyasu Medal and the 2009. 
\end{biography}

\epsfysize=3.2cm
\begin{biography}{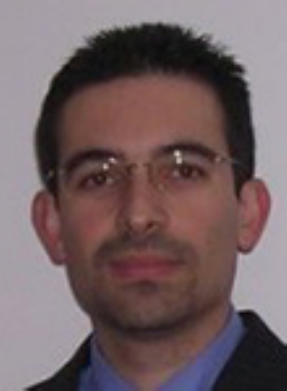}{Carlos T. Calafate}
(calafate@disca.upv.es) graduated with honors in Electrical and
Computer Engineering at the University of Oporto (Portugal) in 2001.
He received his Ph.D. degree in Computer Engineering from the Technical
University of Valencia in 2006, where he has worked since 2005. He
is currently an associate professor in the Department of Computer
Engineering at the UPV, Spain. He is a member of the Computer Networks
research group (GRC). His research interests include mobile and pervasive
computing, security, and quality of service in wireless. 
\end{biography}

\epsfysize=3.2cm
\begin{biography}{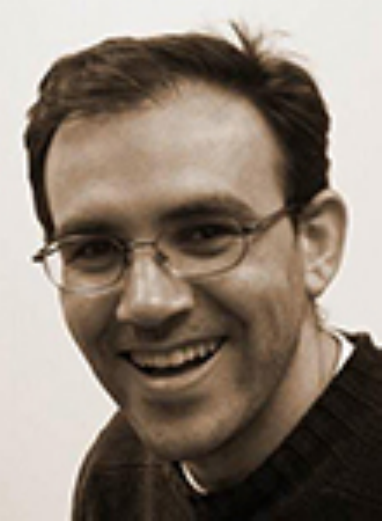}{Juan-Carlos Cano}(jucano@disca.upv.es) received his M.Sc. and Ph.D. degrees in computer
science from the Polytechnic University of Valencia (UPV) in 1994
and 2002, respectively. From 1995 to 1997 he worked as a programming
analyst at IBM\textquoteright s manufacturing division in Valencia.
He is a full professor in the Department of Computer Engineering at
UPV in Spain. His current research interests include power-aware routing
protocols and quality of service for mobile ad hoc networks and pervasive
computing.
\end{biography}

\epsfysize=3.2cm
\begin{biography}{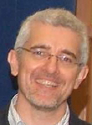}{Pietro Manzoni}(pmanzoni@disca.upv.es) received his M.S. degree in computer science
from the Universita degli Studi of Milan, Italy, in 1989 and the Ph.D.
degree in computer science from the Polytechnic University of Milan,
Italy, in 1995. He is a full professor in the Department of Computer
Engineering at UPV in Spain. His research activity is related to wireless
networks protocol design, modeling, and implementation.
\end{biography}

\end{document}